  \documentclass[prl,twocolumn,superscriptaddress,showpacs]{revtex4}
  \usepackage{graphicx}
  \usepackage{amsmath}
 
\def\beq {\begin{equation}}
\def\eeq {\end{equation}}

\def\beqn  {\begin{displaymath}}
\def\eeqn {\end{displaymath}}

\def\beqarr {\begin{multline}}
\def\eeqarr {\end{multline}}

\def\bv {{\mid}}

\def\w {\omega}

\def\bfk {\mathbf{k}}
\def\bfr {\mathbf{r}}
\def\bfq {\mathbf{q}}

\newcommand{\bra}[1]{\langle #1|}
\newcommand{\ket}[1]{|#1\rangle}

\begin{document}
  \title{Sodium: a charge-transfer insulator at high pressures}
  \author{Matteo Gatti}
   \email{matteo.gatti@ehu.es}
  \affiliation{Nano-Bio Spectroscopy group and ETSF Scientific Development Centre, 
  Dpto. F\'isica de Materiales, Universidad del Pa\'is Vasco, 
  Centro de F\'isica de Materiales CSIC-UPV/EHU-MPC and DIPC, 
  Av. Tolosa 72, E-20018 San Sebasti\'an, Spain}

  \author{Ilya V. Tokatly}
  \affiliation{Nano-Bio Spectroscopy group and ETSF Scientific Development Centre, 
  Dpto. F\'isica de Materiales, Universidad del Pa\'is Vasco, 
  Centro de F\'isica de Materiales CSIC-UPV/EHU-MPC and DIPC, 
  Av. Tolosa 72, E-20018 San Sebasti\'an, Spain}
  \affiliation{IKERBASQUE, Basque Foundation for Science, E-48011 Bilbao, Spain}

  \author{Angel Rubio}
    \affiliation{Nano-Bio Spectroscopy group and ETSF Scientific Development Centre, 
  Dpto. F\'isica de Materiales, Universidad del Pa\'is Vasco, 
  Centro de F\'isica de Materiales CSIC-UPV/EHU-MPC and DIPC, 
  Av. Tolosa 72, E-20018 San Sebasti\'an, Spain}
  \affiliation{Fritz-Haber-Institut der Max-Planck-Gesellschaft, 
  Theory Department, Faradayweg 4-6, D-14195 Berlin-Dahlem, Germany}

 \date{\today}

\begin{abstract}
By means of first-principles methods we analyse the optical response of transparent dense sodium  
as a function of applied pressure. We discover an unusual kind of charge-transfer exciton that
proceeds from the interstitial distribution of valence electrons repelled away from
the ionic cores by the Coulomb interaction and the Pauli repulsion.
The predicted absorption spectrum shows a strong 
anisotropy with light polarization that just at pressures above the
 metal-insulator transition manifests as sodium being optically transparent in one direction
but reflective in the other. This result provides a key information about
the crystal structure of transparent sodium, a new unconventional inorganic electride.
\end{abstract}

  \maketitle
 
  %%%%%%%%%%%%%%%%%%%%%%%%%%%%%%%%%%%%%%%%%%%%%%%%%%%%%%%%%%%%%%%%%%
  %%%%%%%%%%%%%%%%%%%%%% BODY OF THE PAPER %%%%%%%%%%%%%%%%%%%%%%%%%
  %%%%%%%%%%%%%%%%%%%%%%%%%%%%%%%%%%%%%%%%%%%%%%%%%%%%%%%%%%%%%%%%%%

{\it Simple-to-complex  
phase transition.}-- 
In textbooks sodium is defined 
as the simplest of the simple metals \cite{grosso}. 
At ambient pressure it has a high-symmetry crystal structure with non directional metallic bondings.
Valence electrons can be treated as almost independent particles, only weakly perturbed by
the electron-ion interaction.
They form a homogeneous and isotropic electron gas, 
giving rise to a nearly spherical Fermi surface.
As a consequence, also the optical properties are well described by the Drude model of free carriers.
Application of pressure, instead, changes the situation quite dramatically while revealing an emergent complexity.
Sodium under pressure shows an unexpected and often counter-intuitive behaviour. 
Instead of  becoming more free-electron-like, at higher densities it adopts several
low symmetry (and even incommensurate) crystal structures \cite{structure1,structure2,melting1,incommensurate,nature,pnas}
and has a very unusual melting curve \cite{melting1,melting2}.
This surprising behaviour had been anticipated by Neaton and Ashcroft \cite{ashcroft-neaton},
who also predicted a possible metal to insulator transition \cite{mit2}.  
At very high pressures compression is sufficiently strong that core electrons start to overlap and Pauli exclusion principle  
favours less symmetric charge distributions with $p$-$d$ hybridizations \cite{nature,ashcroft-neaton}. 
The Coulomb repulsion between core and valence electrons  
leads to charge localization in interstitial positions, instead of around nuclei, and finally to an insulating state \cite{ashcroft-neaton}.
{\it From a simple, uncorrelated, reflective metal, 
sodium turns into a complex, correlated, transparent insulator. It is a whole change of paradigm} \cite{ashcroft2008}.

The crystal structure of the recently discovered insulating state 
was identified as a distorted, six-coordinated double-hexagonal closed-packed structure (hP4) \cite{nature}.
Experimentally, it is found that under pressure  body-centred cubic  Na first becomes face-centred cubic (at 65 GPa) and then transforms into a cI16 structure \cite{structure1} (at 103 GPa). Besides adopting many different structures near the minimum of the melting curve \cite{structure2} (118 GPa), it crystallizes in the oP8 and tI19 phases \cite{incommensurate} (above 125 GPa), 
before undergoing a final transition to the newly proposed hP4 phase at around 200 GPa. 
Theoretically \cite{nature}, the incommensurate tI19 and the oP8 phases were found to have nearly equal enthalpies and the transition to the hP4 structure occurs at slightly higher pressures (260 GPa). 
Measurements were performed at room temperature, while the zero-temperature structural calculations neglected entropic contributions 
and lattice dynamics (phonon enthalpies along with quantum effects on the light sodium ions), 
and hence could not provide a direct confirmation of the theoretical results \cite{ashcroft_comment}. 

In the present Letter, we study the optical response of the hP4 phase
employing the highly accurate Bethe-Salpeter equation (BSE), 
which is the state-of-the-art approach for first-principles  
predictions of optical properties in solids \cite{RMP-noi}.
Our results provide evidence for the emergence of relevant excitonic (i.e. electron-hole correlation) effects, 
with the formation of an unusual charge-transfer exciton, and a  strong anisotropy in the dielectric response.
They represent a stringent fingerprint of the hP4 crystal structure of transparent dense sodium.

{\it Absorption spectra and excitons.}-- 
The BSE can be cast into an effective two-particle Schr\"odinger  equation 
for the wavefunction $\Psi_\lambda(\bfr_h,\bfr_e)$ 
of the electron-hole pair (i.e. the exciton):
\beq
H_{exc}\Psi_\lambda = E_\lambda \Psi_\lambda,
\label{eqBSE}
\eeq
where the excitonic Hamiltonian $H_{exc}$ includes the electron-hole interaction \cite{RMP-noi}.
In our first-principles approach, $\Psi_\lambda$ is represented on a Kohn-Sham basis:
\beq
\Psi_\lambda(\bfr_h,\bfr_e) = \sum_{{\bfk}vc}  \Psi_\lambda^{{\bfk}vc} \phi_{v\bfk}^*(\bfr_h)\phi_{c\bfk}(\bfr_e),
\eeq
where $v$ ($c$) runs on valence (conduction) bands and $\bfk$ is in the first Brillouin zone.
The optical spectrum is then obtained from the imaginary part of the dielectric function $\epsilon_2 = \text{Im} \, \epsilon$:
\beq
\epsilon_2(\w) =  \lim_{\mathbf{q} \rightarrow 0} \frac{8\pi}{q^2} \sum_\lambda \Big|\sum_{{\bfk}vc} \Psi_\lambda^{{\bfk}vc} 
\bra{\phi_{v\bfk+\bfq}}e^{-i\mathbf{q}\mathbf{r}}\ket{\phi_{c\bfk}} \Big|^2 \delta(\w-E_\lambda).
\label{spectrumBSE}
\eeq
When excitonic effects are unimportant
the eigenvalues $E_\lambda$ correspond to vertical transition between 
one-quasiparticle energies $E_{c\bfk} - E_{v\bfk}$ (which we calculate in the GW approximation \cite{Hedin}) 
and the coefficients $\Psi_\lambda^{{\bfk}vc}$  become diagonal, reducing Eq. \eqref{spectrumBSE} 
to the well known Fermi's golden rule \cite{grosso}.
We investigate the effect of pressure on the formation of excitons,
through charge localization and the reduction of the screening of the electron-hole interaction.
Excitons can be identified by means of the twofold role that they play on the absorption spectra.
First, the modification of the excitation energies induces a shift of the positions of the peaks in the spectrum: 
as a consequence, new peaks can appear inside the fundamental band gap (bound excitons);
second, the mixing of the independent-particle transitions leads to a global reshaping of the spectrum towards lower energies (continuum excitons).
Our {\it ab initio} solution of the BSE proceeds in three steps. 
We start from a ground-state calculation using density-functional theory in local-density approximation (LDA), 
where we employ the  lattice constants of the hP4 phase
from Ref. \cite{nature} corresponding to representative pressures (see Ref. \cite{epaps} Tab. 1). 
We treat 3$s$, 2$p$ and 2$s$ electrons explicitly as valence, leaving only the very tightly bound 1$s$ in the frozen ionic core 
(this required us to use a energy cut-off of 80 Ha in the plane-wave representation of the wave functions).
We then calculate the quasiparticle GW electronic structure that we use as input for the Bethe-Salpeter equation \eqref{eqBSE}. 
In the calculation of the screening and the GW corrections, where we adopt the Godby-Needs plasmon-pole model \cite{Godby-Needs},  
we use 150 bands and a cut-off of 20 Ha on the wave functions. 
To converge the absorption spectra up to 7-8 eV we consider 2 valence and 8 conduction bands 
and sample the first Brillouin zone with a uniform grid of inequivalent 2000 $\bfk$ points,
slightly shifted from high-symmetry $\bfk$ points. We then apply a 0.1 eV gaussian broadening \footnote{
We use 1024 $\bfk$ points and 0.3 eV broadening when we study the dependence on pressure of the optical spectra.
For the plot of the exciton wave function the $\bfk$ point grid includes the $\Gamma$ point.}.

\begin{figure}[t]
\begin{center}
\includegraphics[width=\columnwidth]{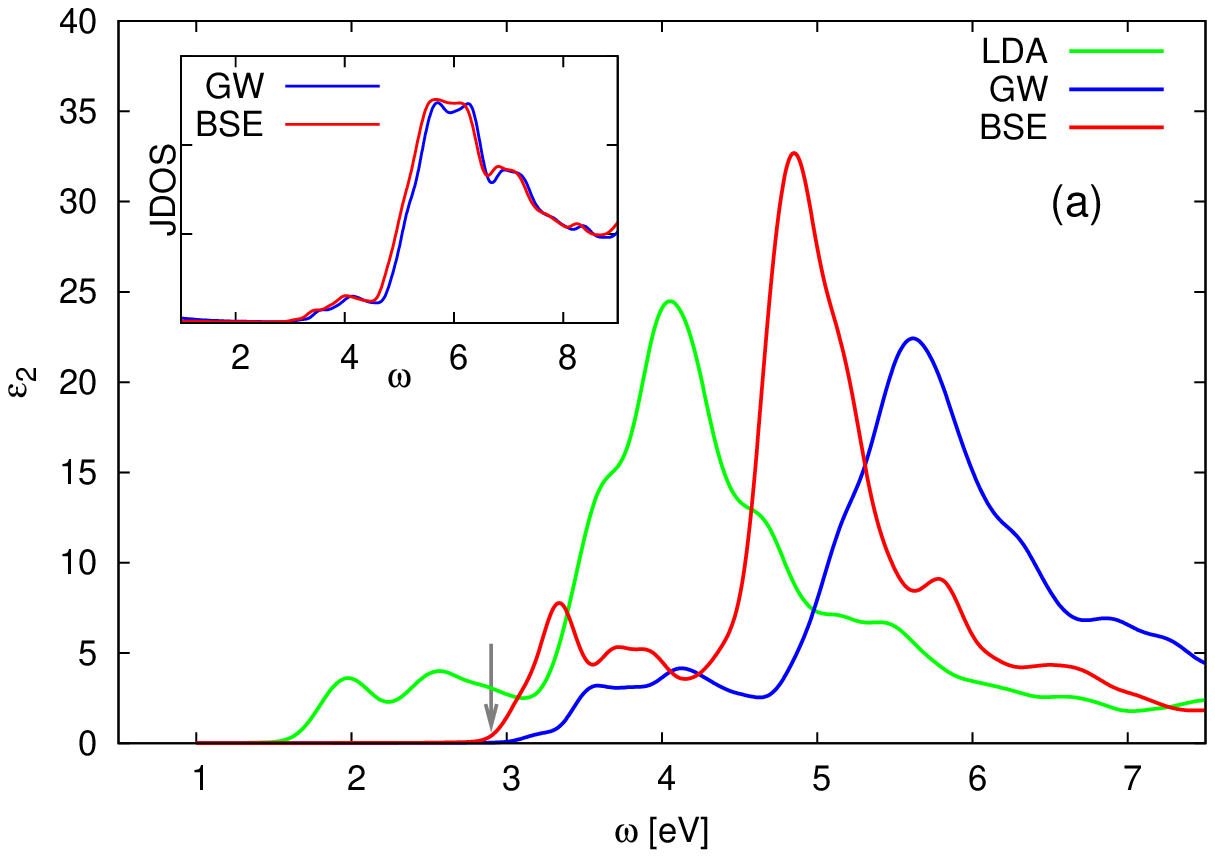} \\[0.25cm]
\includegraphics[width=0.49\columnwidth]{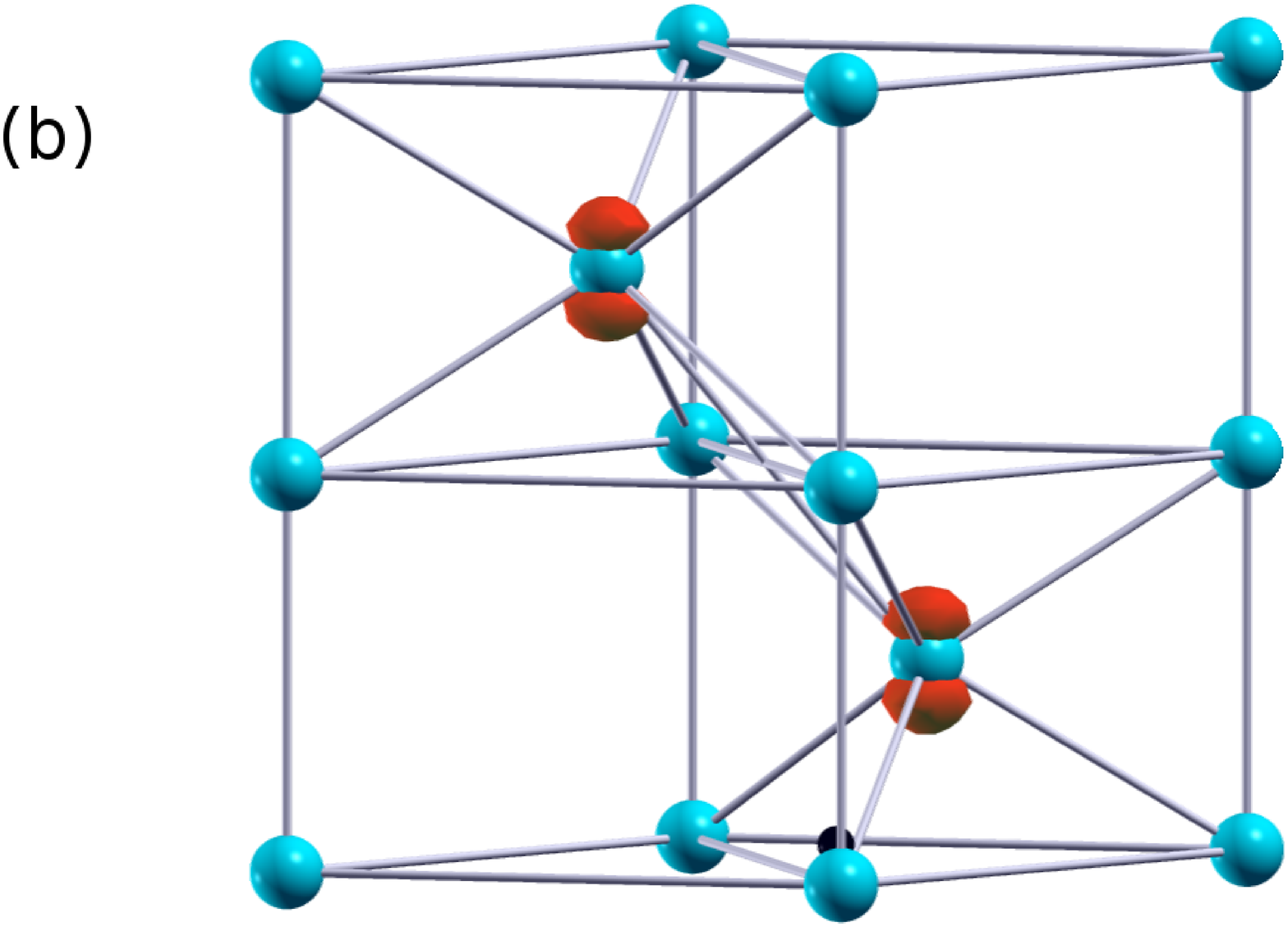} \includegraphics[width=0.49\columnwidth]{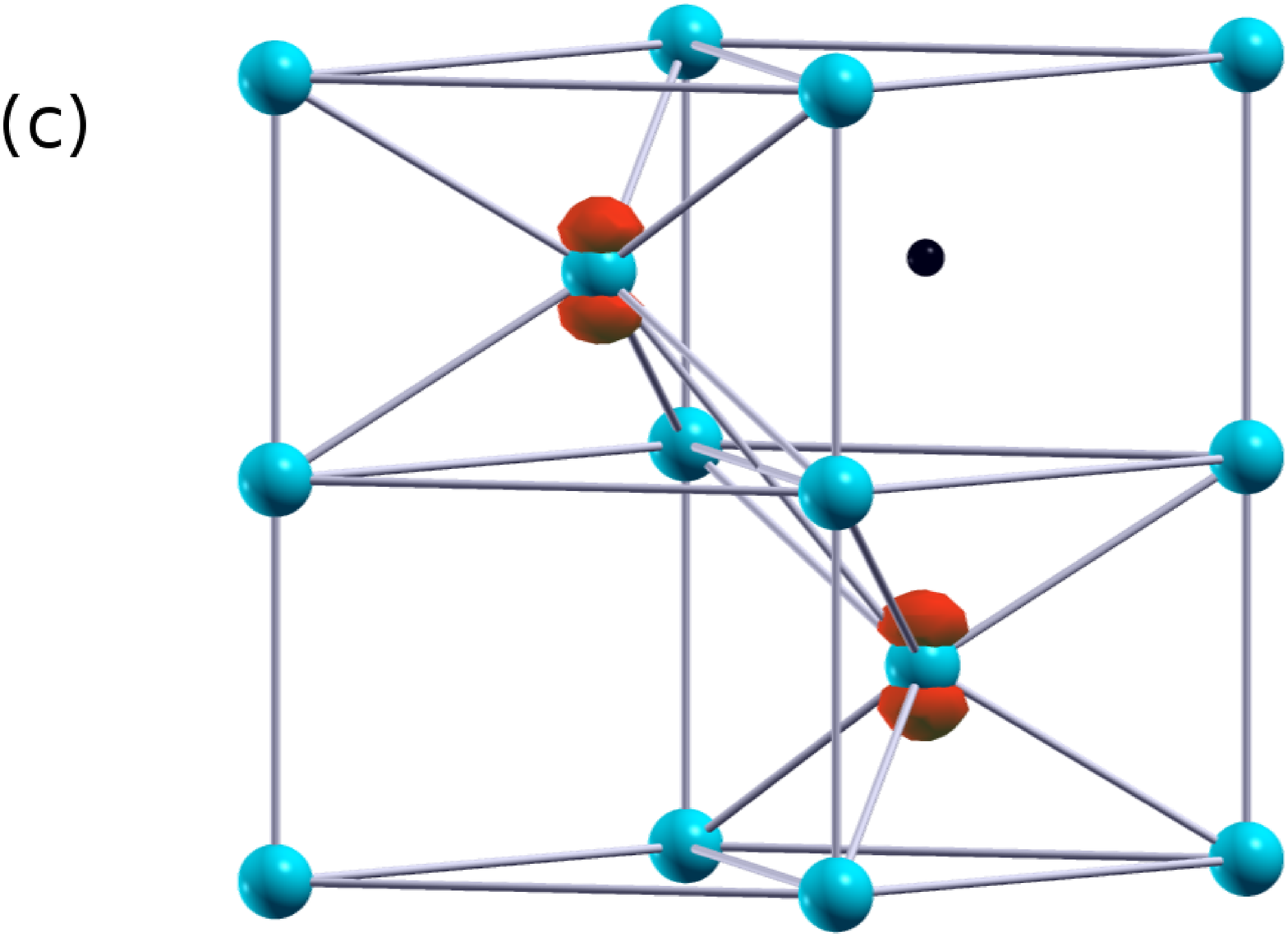}\\[0.1cm]
\includegraphics[width=0.49\columnwidth]{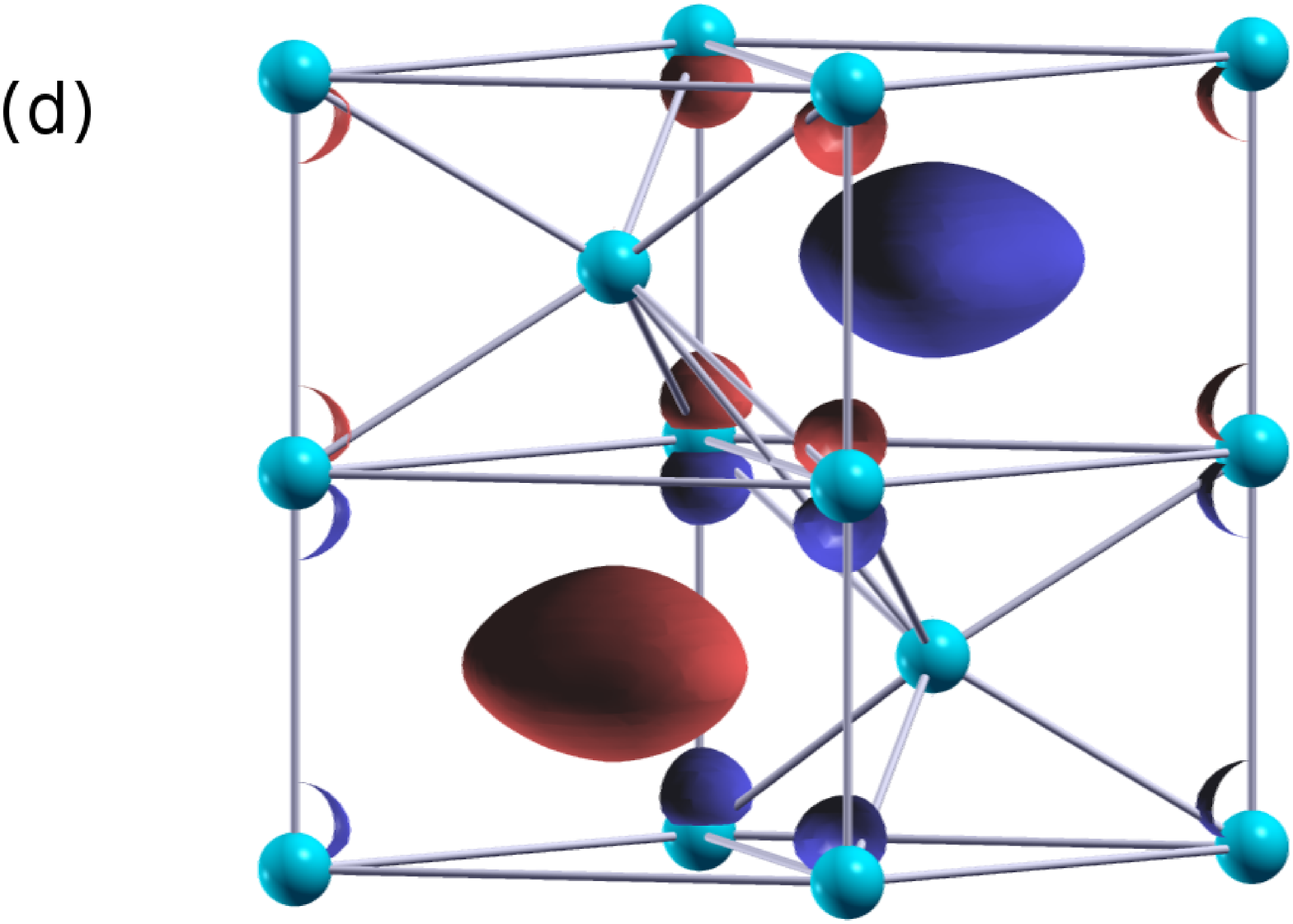} \includegraphics[width=0.49\columnwidth]{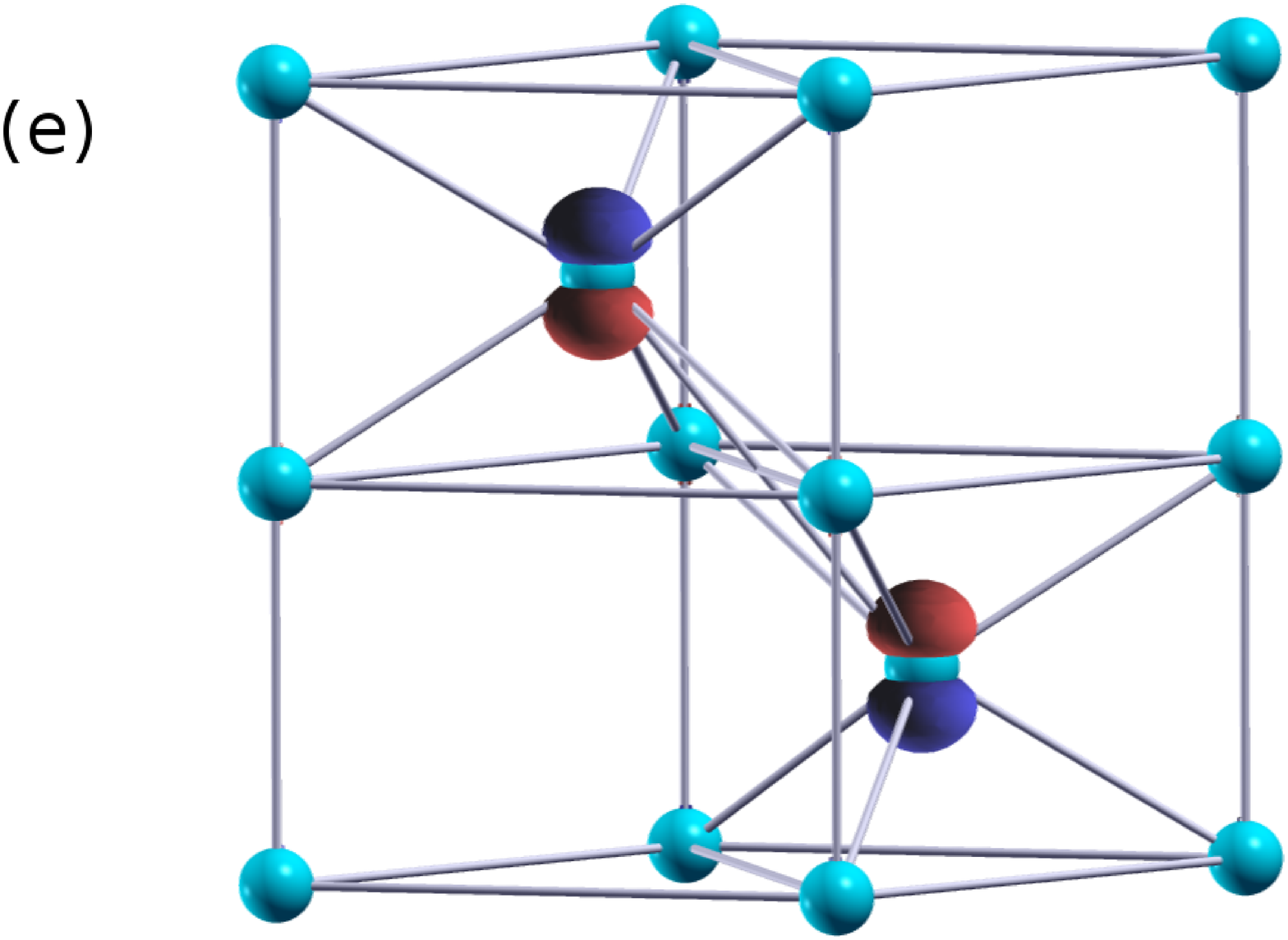}
\caption{(Color online) (a) Excitons in BSE (red curve, main panel) change substantially the shape of the GW (blue) and LDA (green) spectra calculated without electron-hole interactions.  They redshift the absorption onset (grey arrow) by $\sim$0.1 eV with respect to the fundamental GW quasiparticle gap. The apparent shift of the spectrum is only partially accounted for by the change of the JDOS, from the GW excitation energies to the density of excitonic states of the BSE solution (inset). The remaining effect is due to mixing of formerly independent valence-conduction transitions.
(b)-(c) Electron-hole pair wavefunction $\Psi_\lambda(\bfr_h,\bfr_e)$ for the first exciton.
When the hole (black dot) is located either near a corner atom (b) or in interstitial position (c), the electron charge is always on the central atoms (isocontour in red: plot of $\bv\Psi_\lambda\bv^2$ at fixed $\bfr_h$ at 0.3 of its maximum). (d)-(e) Kohn-Sham one-particle real wave functions $\phi_{n\bfk}(\bfr)$, top valence (d) and bottom conduction (e) states at the $\Gamma$ point: isocontours at $+0.3$ (red) and $-0.3$ (blue) of their maximum.}
\label{fig1}
\end{center}
\end{figure}

{\it Excitons in dense sodium.}-- 
To illustrate our discussion,
we first consider the results obtained with the lattice parameters calculated by Ma {\it et al.} \cite{nature}
in correspondence to a pressure of 320 GPa.
We then analyse the dependence of the excitonic effects with applied pressure.
In Fig. \ref{fig1}(a) we compare the absorption spectra that we obtain by solving the BSE or by using Fermi's golden rule
(averaged with respect to the different light polarizations). 
In the latter case transition energies are calculated either using LDA or the GW approximation. 
The GW corrections on the band structure induce an almost rigid blue-shift to the LDA spectrum, 
reflecting the correction of the usual underestimation of the quasiparticle band gap in LDA (see Ref. \cite{epaps} Fig. 1).
The onset moves from 1.7 eV in LDA to 3.0 eV in GW. It corresponds to a direct transition in $\Gamma$ 
between the top valence and the bottom conduction bands.
Adding interaction between GW-quasiparticles, i.e. quasiparticle electron-hole attraction, we observe a substantial modification
of the spectrum: i) strong redistribution of the optical oscillator strength to lower energies, changing the global shape of the spectrum; 
ii) strong narrowing of the main peak at $\sim$ 5~eV and iii) appearance of bound excitons inside the gap.
It turns out that electron-hole attraction reduces the transition energies by about 0.1 eV, determining a global red-shift of the joint density of states (JDOS)  (inset of Fig. \ref{fig1}(a)). 
The absorption onset is now at 2.9 eV (grey arrow in Fig. \ref{fig1}(a)). 
This implies the formation of a bound exciton in the fundamental gap with a binding energy of about 100 meV, that is larger than the one
obtained in standard semiconductors \cite{RMP-noi}.

In Fig. \ref{fig1}(b)-(c) we analyse the shape of the wave function corresponding to this bound electron-hole pair. 
We fix the position of the hole in a particular position $\bfr_h$
and determine the corresponding density distribution of the electron.
Our calculation reveals that this 
bound exciton is, in particular, a charge-transfer exciton.
While the hole is created in the vicinity of one corner atom (Fig. \ref{fig1}(b)), the excitation has largely moved 
the electron charge to localize on the central atoms, forming a separate sub-lattice  over many unit cells 
(see Ref. \cite{epaps} Fig. 2). Remarkably, 
the same electronic distribution is obtained even when the hole is located in the interstitial regions (Fig. \ref{fig1}(c)), 
where the valence electrons accumulate when they are forced away from the atoms \cite{nature,ashcroft2008}.
Charge-transfer excitons are typical of ionic insulators like alkali-halide crystals \cite{abbamonte}.
They are intermediate between strongly bound Frenkel and weakly bound Wannier excitons, as they involve the transfer of one electron 
from a negative ion to a positive ion.
In the present case, in particular, the role of the anion is played by the interstitially confined electron gas. {\it In this sense, 
transparent sodium behaves as an unconventional inorganic electride \cite{dye}, characterised by the simplest possible anion: 
an electronic charge without ionic core.}
The strong enhancement of the peak at around 5 eV, which is the mark of the formation of a resonant exciton, instead
is not explained by the simple modification of the transition energies, since the JDOS curves remain very similar. 
It is rather due to the mixing of a very large number of independent electron-hole transitions in the same energy range 
(see Ref. \cite{epaps} Fig. 1), 
each of them participating with a small contribution to the excitonic peaks. 
It is the sign of highly correlated excitonic states. 

\begin{figure}[t]
\begin{center}
\includegraphics[width=\columnwidth]{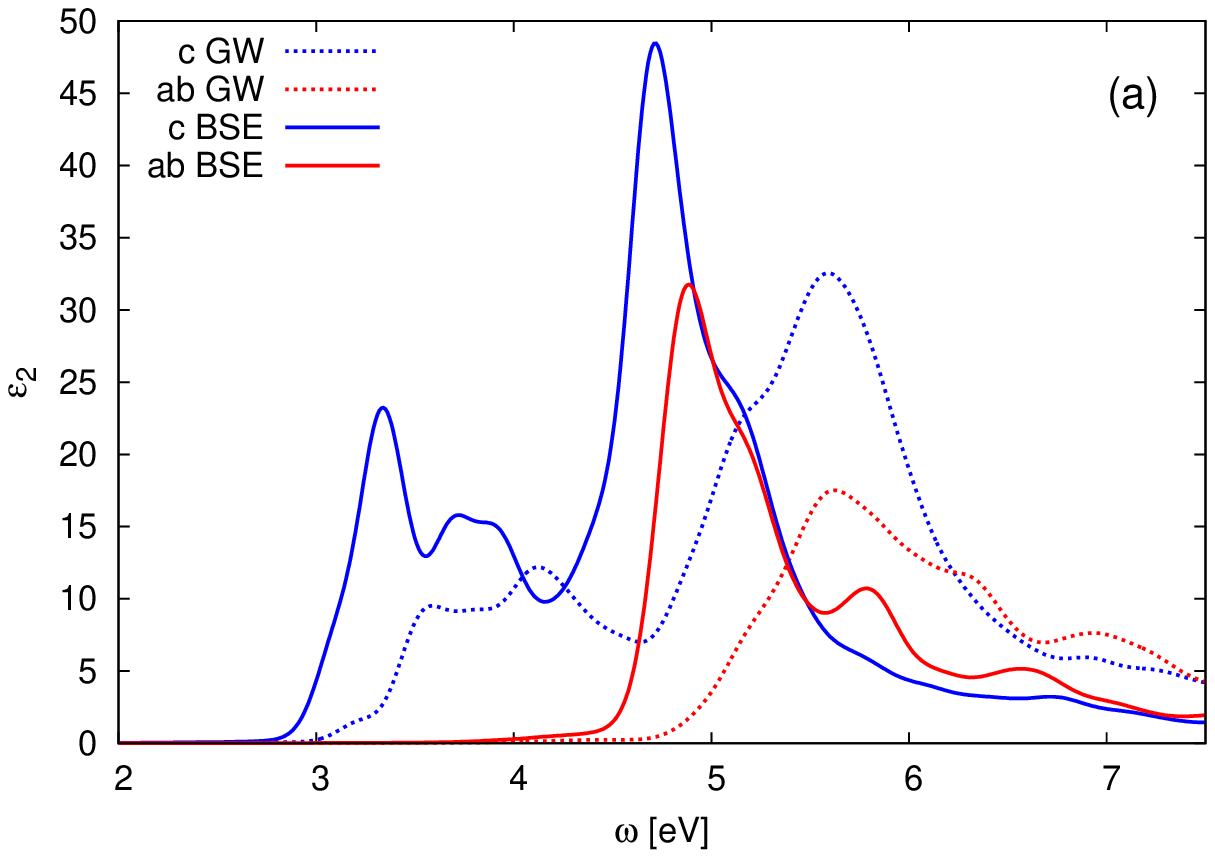} \\[0.1cm]
\includegraphics[width=\columnwidth]{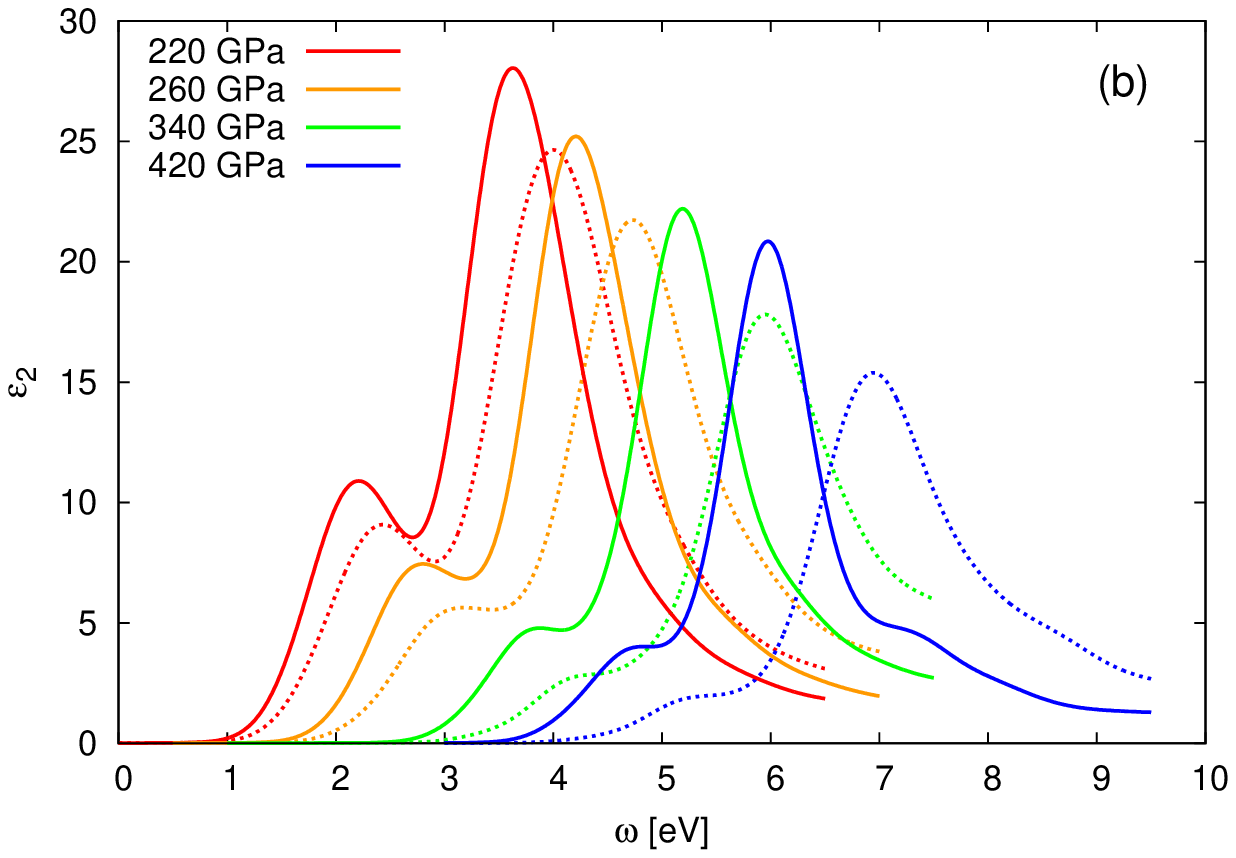} 
\caption{(Color online) (a) Spectra calculated with polarization either along the hexagonal $c$ axis (blue) or in the in-plane $ab$ directions (red). In both cases, dotted lines are for GW spectra without electron-hole interaction and solid lines for BSE spectra with electron-hole interaction. (b) Absorption spectra calculated at lattice parameters corresponding to increasing pressures (data from Ref. \protect\cite{nature}), with or without excitonic effects (solid and dotted lines, respectively).}
\label{fig2}
\end{center}
\end{figure}

{\it Anisotropy and inhomogeneity.}--
By comparing the optical spectra obtained 
with light polarized either along the hexagonal $c$ axis  or in the perpendicular $ab$ plane (Fig. \ref{fig2}(a)), 
we find that the spectra with the two light polarizations remain very different, both with and without the inclusion of excitonic effects.
In particular, in the in-plane direction, sodium is completely transparent up to 4.5 eV. 
In fact, the absorption onset and the first structures in the spectrum in Fig. \ref{fig1}(a) are entirely due to absorption 
of light polarized along the $c$ axis. Also the exciton represented in Fig. \ref{fig1}(b)-\ref{fig1}(c) is dark in the in-plane direction.
This behaviour can be understood as due to the particular symmetry of the one-particle wave functions involved in the transitions.
For instance, by symmetry at the $\Gamma$ point the transition between the top valence (Fig. \ref{fig1}(d))
and the bottom conduction (Fig. \ref{fig1}(e)) has zero oscillator strength in the in-plane direction.
On the other hand, inhomogeneities in the charge distributions are responsible 
for the so-called crystal local-field effects \cite{adler-wiser}.
In general, an external field can induce spatial charge fluctuations that are rapidly varying on the microscopic scale. 
It is intuitively clear that the self-consistent response to these induced local fields becomes more important 
the more the system is inhomogeneous and polarizable. 
Whereas in the absorption spectra of bulk solids local fields are generally weak \cite{RMP-noi},
in dense sodium they induce a sizeable effect (see Ref. \cite{epaps} Fig. 3), particularly for the main peak and $c$ axis polarization (while the onset is unaffected).
This is a further indication that the charge distribution has become very inhomogeneous.

\begin{figure}[t]
\begin{center}
\includegraphics[width=\columnwidth]{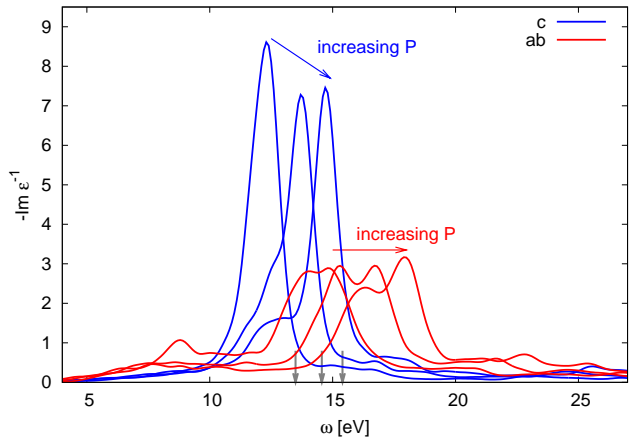} 
\caption{(Color online) EELS spectra, $-\text{Im} \, \epsilon^{-1}(\bfq,\w)$, along the two polarizations for a momentum transfer $\bfq$ equal to 0.1 reciprocal lattice units at 3 different pressures: 220, 320, and 420 GPa.  The grey arrows mark the plasmon energies obtained using a Drude model for a homogeneous electron gas with a density corresponding to 4 valence electrons per unit cell.}
\label{fig3}
\end{center}
\end{figure}

{\it Dependence on the pressure.}-- 
The application of increasing pressure  induces as a first effect a band gap opening in the quasiparticle band structure \cite{nature}.
As a consequence also the optical spectra move to higher energies (Fig. \ref{fig2}(b)).
Moreover, the quasiparticle gap from indirect becomes direct at the $\Gamma$ point (see Ref. \cite{epaps} Fig. 5).
Additional compression leads to localize more the electronic charge and to reduce the screening of the interaction 
between the electron and the hole. 
Both contribute to increase the excitonic effects: the electron and the hole on average stay closer 
and their effective interaction is stronger.
This explains why the largest differences between 
the spectra with and without electron-hole interaction occur for the spectrum at the highest pressure (Fig. \ref{fig2}(b)). 
In fact, the static dielectric constant, which we calculate in random-phase approximation (RPA) \cite{RMP-noi}, decreases with increasing pressure, 
passing from 18 at 220 GPa to 7 at 420 GPa.
In addition, we consider the dynamical screening, $-\text{Im} \, \epsilon^{-1}(\bfq,\w)$, 
which can be measured by electron-energy loss spectroscopy (EELS) 
or inelastic x ray scattering (IXS) and which we calculated in time-dependent LDA \cite{hansi} (with the inclusion of 40 bands).
The EELS spectrum, for each momentum transfer $\bfq$ (either along the $c$ axis or in the perpendicular $ab$ plane), is characterised by a main structure, 
which corresponds to the frequency of the collective oscillations of the valence electrons in the system, i.e. plasmons  (Fig. \ref{fig3}). 
These plasmon structures show a small dispersion with the  momentum transfer $\bfq$ (see Ref. \cite{epaps} Fig. 6). 
While detailed features depend on band structure properties, in this case the Drude model is still partially able to track the dependence of the plasmon frequency $\omega_p$ with the compression.
According to the Drude model, the plasmon frequency increases linearly with the square root of the density \cite{grosso}: $\omega_p = \sqrt{4\pi \rho}$. 
Using this formula 
 at the densities corresponding to the three considered pressures we have  $\omega_p =$ 13.5, 14.6 and 15.4 eV   (grey arrows in Fig. \ref{fig3}) when we consider only 4 valence electrons in the unit cell. In a way, this corresponds to a definition of ``valence'' electrons in this extreme situation 
where the usual view of electrons belonging to a particular atom is no more strictly valid. 
Being based on a uniform electron gas, the Drude model describes an average behaviour of the dielectric response of the system,
but is obviously not able to catch its anisotropy, 
which leads to the splitting between the plasmons along the $c$ axis 
and in the perpendicular plane (the Drude plasmon energy falls somewhat in the middle). Moreover, at low pressures a smaller further plasmonic structure appears only in the in-plane direction at about 8 eV, which is strongly suppressed by increasing the applied pressure (see Ref. \cite{epaps} Fig. 7).
Both the anisotropy in the lattice structure and the higher-energy localisation of the spectral weight for the in-plane polarization (see Ref. \cite{epaps} Fig. 7) contribute in the same way to the splitting of the plasmon energies in the two directions.

In conclusion, our first-principle predictions for both absorption 
and EELS spectra can be indeed used to confirm the crystal structure of this phase of dense transparent sodium.
First of all, since the binding energy of the bound exciton is of the order of a hundred of meV, 
our results for the optical spectra support the hypothesis of Ma {\it et al.} \cite{nature}, which was based only on ground-state 
and band gap calculations.  
Moreover, even though ground-state calculations overestimate the pressure value at which the transition to the hP4 phase occurs \cite{nature}, 
our predictions for the optical properties are not biased by this error. 
In fact, while excitonic effects are varying with the applied pressure (and can be influenced by the temperature as well \cite{andrea}), 
from our analysis it turns out that key features of the spectra like the polarization anisotropy
 are essentially linked to the symmetry of the crystal structure and can be observed in a very large range of pressure (see Ref. \cite{epaps} Fig. 4). 
This also implies that sodium becomes transparent in the visible at a lower pressure in one polarization direction 
while being still ``metallic'' in the other.
Therefore we suggest the measurement of the spectra along the two polarizations
as the key test to verify conclusively that transparent sodium crystallizes in the hP4 structure.
Photoluminescence would put in evidence the nature of the bound exciton predicted in the present work and that has a low
oscillator strength in optics.

 We acknowledge fruitful discussions with  G. Vignale and F. Sottile 
and the authors of Ref. \cite{nature} for sending their relaxed
geometries used in the present work.
This work was supported by the Spanish MEC (FIS2007-65702-C02-01),
``Grupos Consolidados UPV/EHU del Gobierno Vasco" (IT-319-07), 
the European Union through  e-I3 ETSF project (Contract Number 211956). We acknowledge 
support by the Barcelona Supercomputing Center, ``Red Espanola de Supercomputacion".
We have used Abinit \cite{abinit}, Dp and Exc codes \cite{dp-exc}.

 %%%%%%%%%%%%%%%%%%%%%% REFERENCES %%%%%%%%%%%%%%%%%%%%%%%%%

\begin{widetext}

\clearpage
\setcounter{figure}{0}

\begin{center}
{\Large {\bf  Supplementary information}}
\end{center}

\vspace{5.0cm}

\begin{table*}[ht]
\begin{center}
\begin{tabular}{| c |c |c|}
\hline
a & c & P \\
\hline
2.874 & 4.238 & 220 \\
2.838 & 4.057 & 260 \\
2.769 & 3.817 & 340 \\
2.705 & 3.668 & 420\\
\hline
\end{tabular}
\caption{The lattice parameters of the Na-hP4 phase used in the present work (in \AA) at different pressures (in GPa), from Y. Ma {\it et al.},  Nature  {\bf 458}, 182 (2009).}
\end{center}
\end{table*}

\begin{figure}[ht]
\begin{center}
\includegraphics[width=0.85\columnwidth]{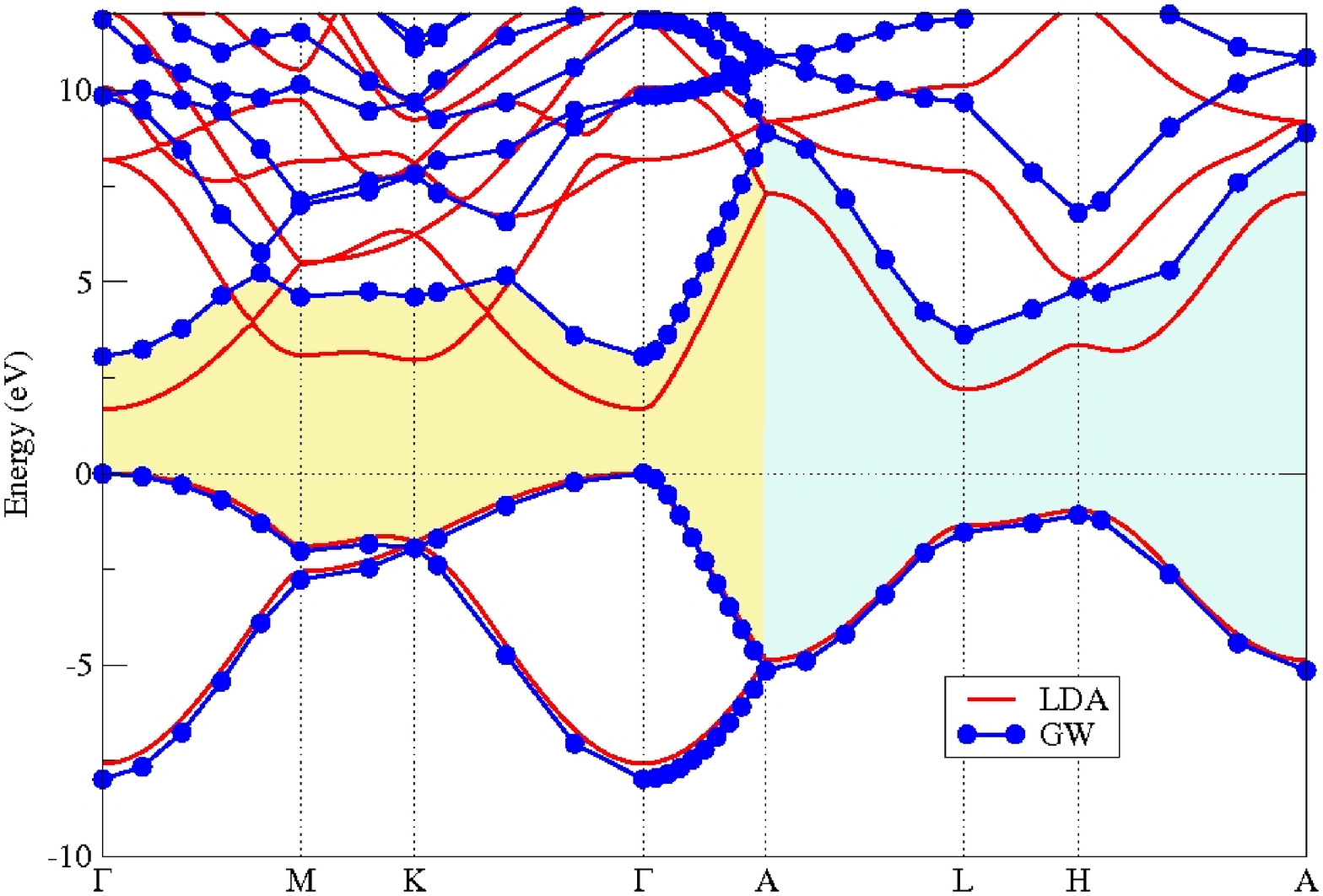} 
\caption{{\bf Band structure at 320 GPa.} The GW corrections (blue) induce an almost rigid upshift of the conduction states calculated in LDA (red).
The zero of the energy scale is set at the top-valence energy. The fundamental gap is direct at $\Gamma$: 1.7 eV in LDA and 3.0 eV in GW.  
In yellow: region of the Brillouin zone where the vertical transitions from top valence to bottom conduction states have zero dipole oscillator  strength in the in-plane direction. They contribute only to the absorption for light polarized along the hexagonal $c$ axis, forming the first peak in the spectrum at lower energies. In light-blue: zone where they contribute to both polarizations (in this case the top valence band is twofold degenerate).}
\label{figgw}
\end{center}
\end{figure}

\begin{figure}[ht]
\begin{center}
\includegraphics[width=0.6\columnwidth]{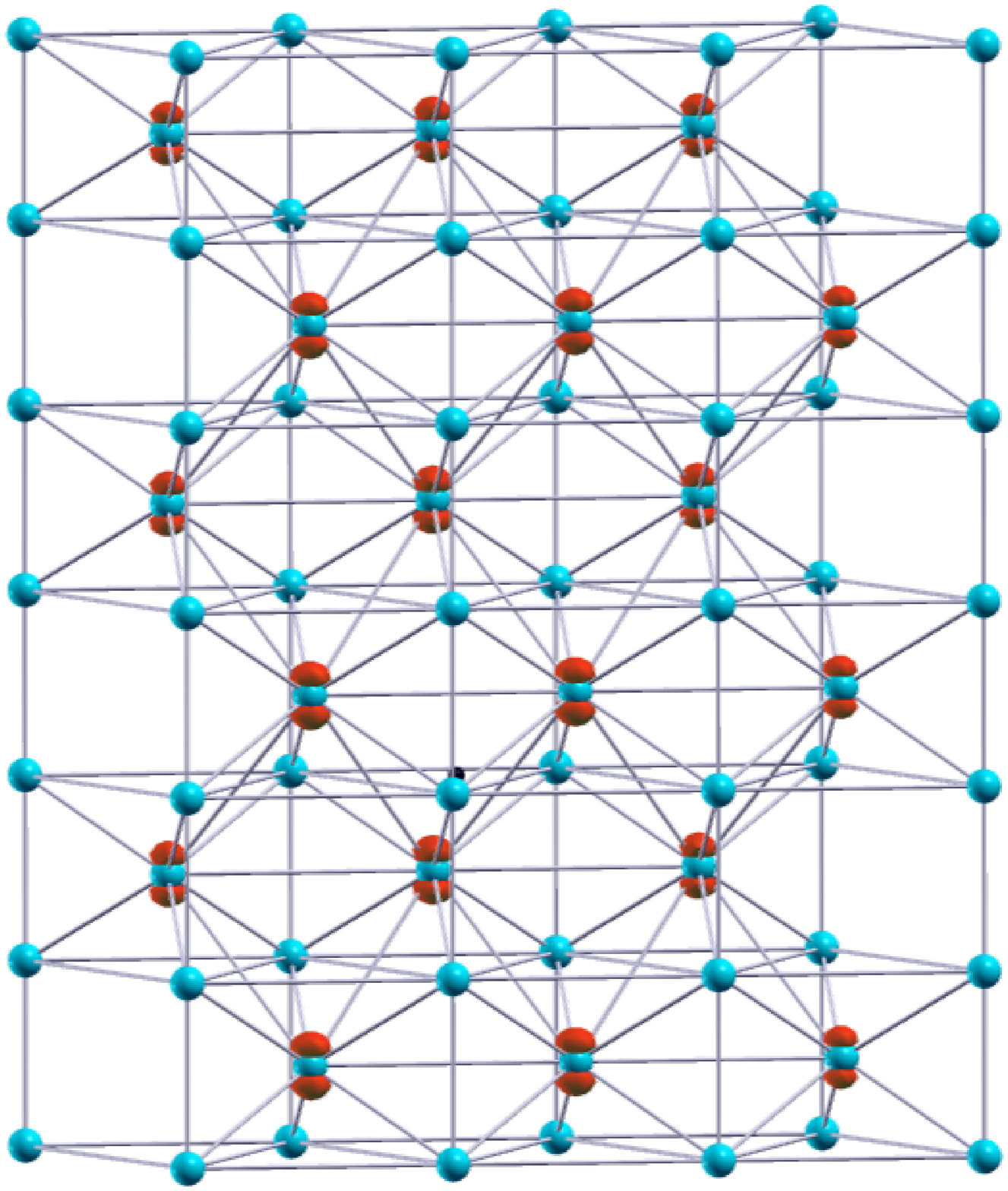} 
\caption{{\bf Exciton wavefunction.} Isocontour plot of the exciton wavefunction $\Psi_\lambda(\bfr_h,\bfr_e)$  at 0.3 of its maximum.
 The position $\bfr_h$ of the hole is fixed on one corner atom (black dot). The isocontour represents the electron charge extending over many unit cells (here shown a 3x1x3 supercell). This implies that the electron and the hole are not strongly bound (as it would be for a Frenkel exciton). Na behaves as a ``bipartite'' system: the hole lies on a one site and the electrons on the others. }
\label{exciton_manycells}
\end{center}
\end{figure}

\begin{figure}[t]
\begin{center}
\includegraphics[width=0.75\columnwidth]{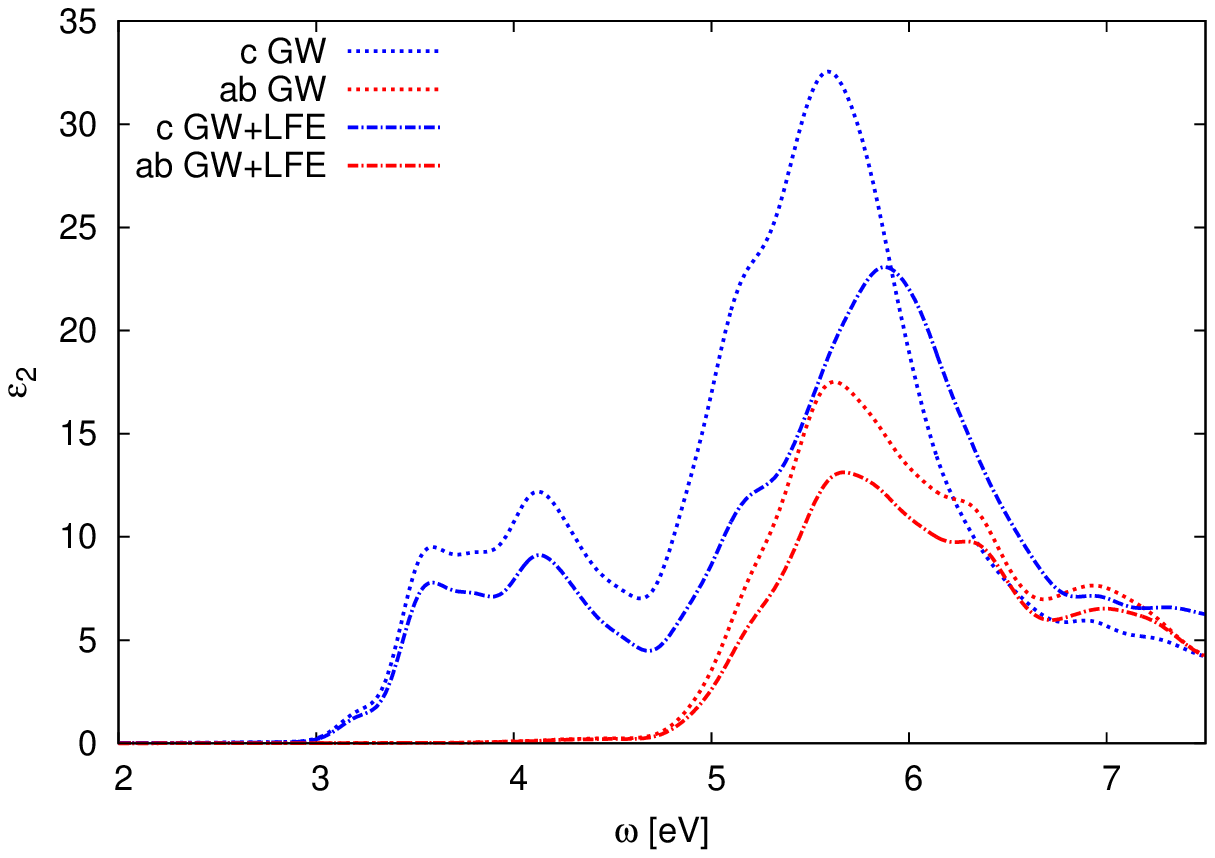} 
\caption{{\bf Crystal local-field effects.} Dotted lines: absorption spectra calculated with Fermi's golden rule using GW quasiparticle energies. 
Dot-dashed lines: with the inclusion of crystal local fields. The effect is quite strong, in particular for the $c$ axis polarization (blue). The spectral weight is blueshifted (but the absorption onset does not change) and its intensity reduced. In $sp$ semiconductors the effect of local fields on absorption spectra normally is less relevant than what found here. Local fields become more important when the system is less homogeneous (but still polarizable). This is a manifestation of the very inhomogeneous distribution of the electronic charge.}
\label{figlfexz}
\end{center}
\end{figure}

\begin{figure}[ht]
\begin{center}
\includegraphics[width=0.75\columnwidth]{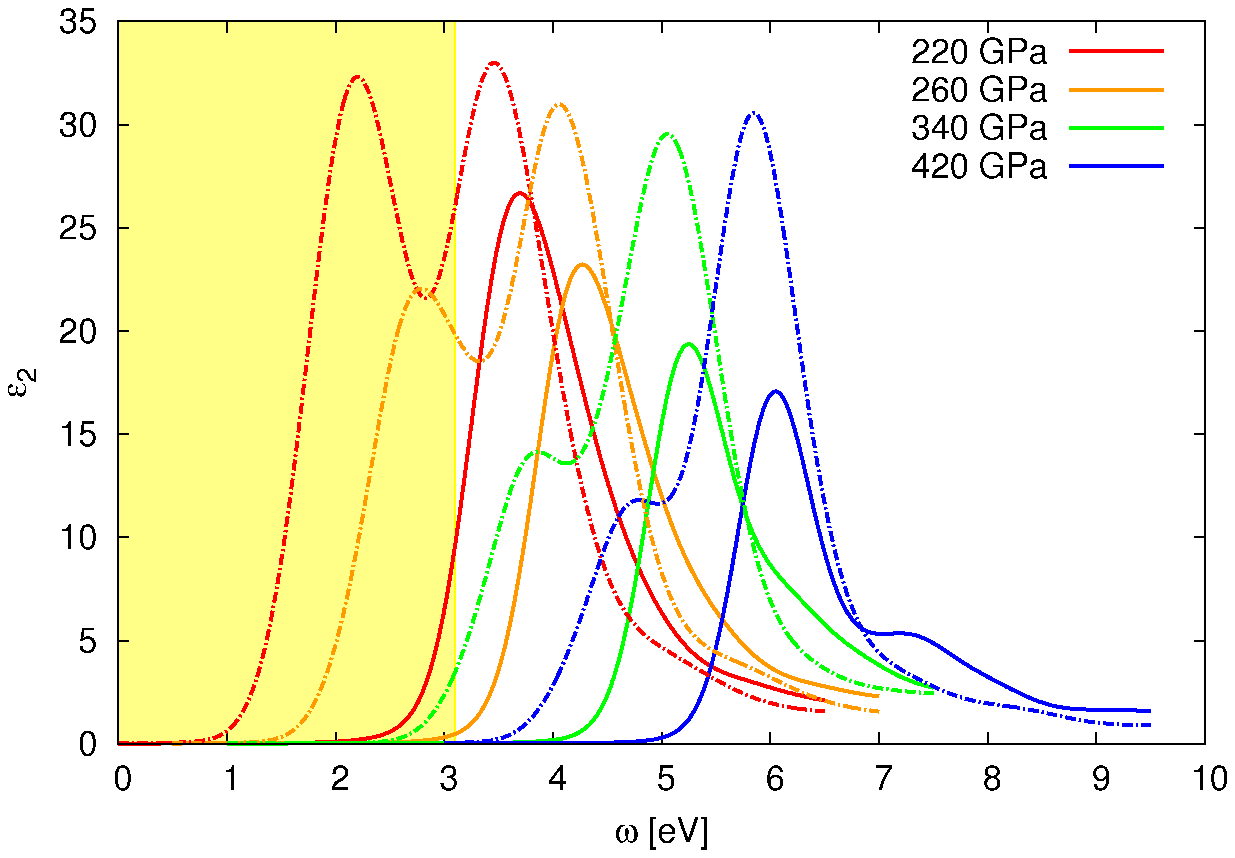} 
\caption{{\bf Polarization dependence of the transparency.} Yellow zone: energy range of visible light (up to 3.1 eV).
At low pressures Na is transparent in the in-plane direction (solid lines) and reflective along $c$ axis (dot-dashed lines).
By increasing pressure, the peak in the in-plane direction (solid lines) is quite rigidly blueshifted and its intensity is reduced accordingly (as an effect of the $f$-sum rule). Instead, for the polarization along the $c$ axis (dot-dashed lines) at low pressures the spectrum is characterised by a double peak, while by increasing pressure the first peak progressively looses strength developing as a shoulder of the second one.} 
\label{figbsep_2}
\end{center}
\end{figure}

\begin{figure}[t]
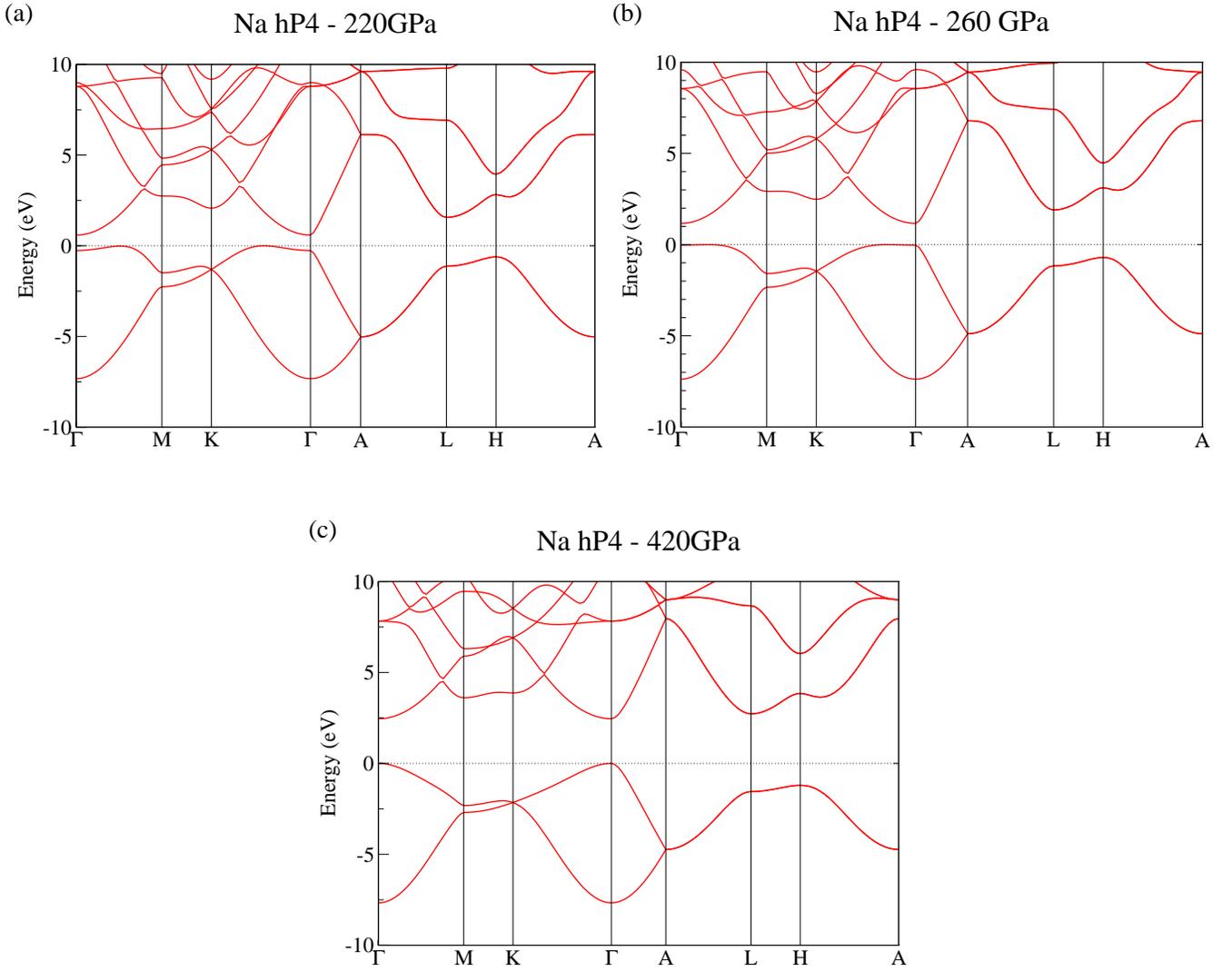

\begin{center}
\includegraphics[width=0.49\columnwidth]{fig_s5a} 
\includegraphics[width=0.49\columnwidth]{fig_s5b} \\[1cm]
\includegraphics[width=0.49\columnwidth]{fig_s5c}
\caption{{\bf From indirect to direct semiconductor.} Evolution of the LDA band structure with applied pressure. At lower pressures Na has an indirect gap (the bottom conduction is at $\Gamma$ and the top valence along $\Gamma-K$ and $\Gamma-M$). By increasing pressure, the gap becomes direct at $\Gamma$. The change occurs at the pressure P=260 GPa where in LDA the hP4 structure becomes energetically more stable than the tI19 one. While the valence bandwidth is quite insensitive to pressure, there appears a clear band gap opening. The zero of the energy scale is located in correspondence to the top valence energy. }
\label{bands220420}
\end{center}
\end{figure}

\begin{figure}[ht]
\begin{center}
\includegraphics[width=0.75\columnwidth]{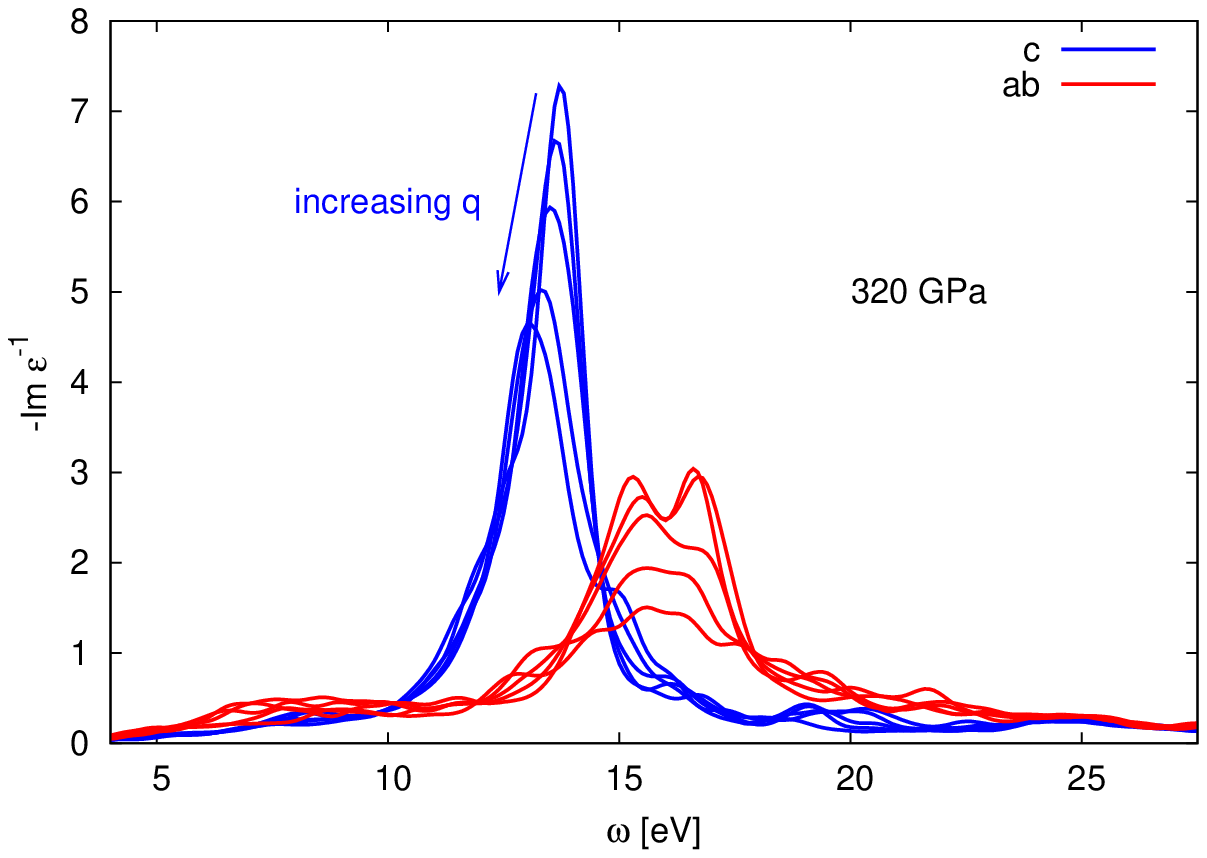} 
\caption{{\bf Plasmon dispersion.} EELS spectra calculated in TDLDA for increasing momentum transfers $\bfq$, either along the $c$ axis (blue) or in the perpendicular $ab$ plane (red). The lattice parameters correspond to a pressure of 320 GPa. In the direction of the arrow, at each curve the momentum transfer is increased by 0.1 reciprocal lattice units, from 0.1 to 0.5. Both the plasmon energy and the intensity of the plasmon peaks reduce with increasing $\bfq$, but the dispersion with $\bfq$ is not very large. The double plasmon peak in the $ab$ plane (red) tends to broaden into a single structure. This means that the polarization anisotropy of the loss function is robust with respect to the momentum transfer.}
\label{figeels_2}
\end{center}
\end{figure}

\begin{figure}[ht]
\begin{center}
\includegraphics[width=0.49\columnwidth]{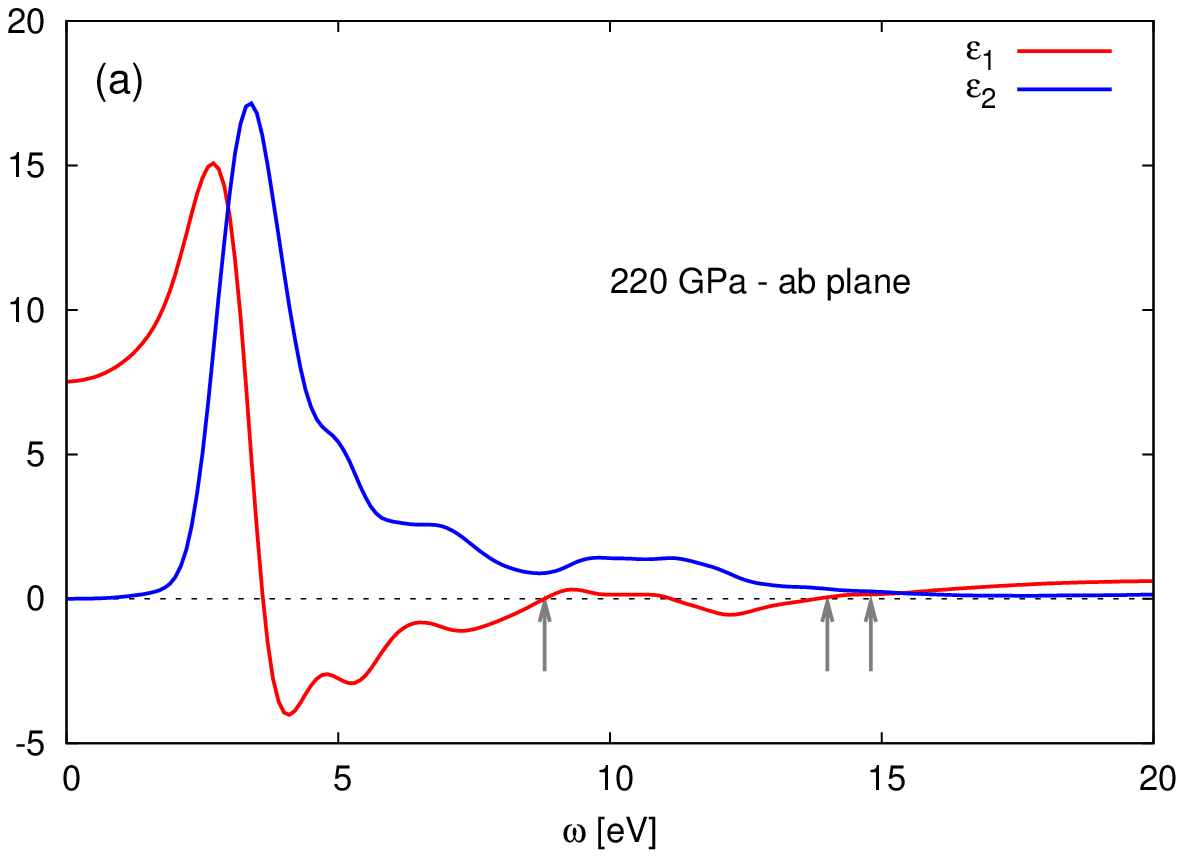} \\[0.66cm]
\includegraphics[width=0.49\columnwidth]{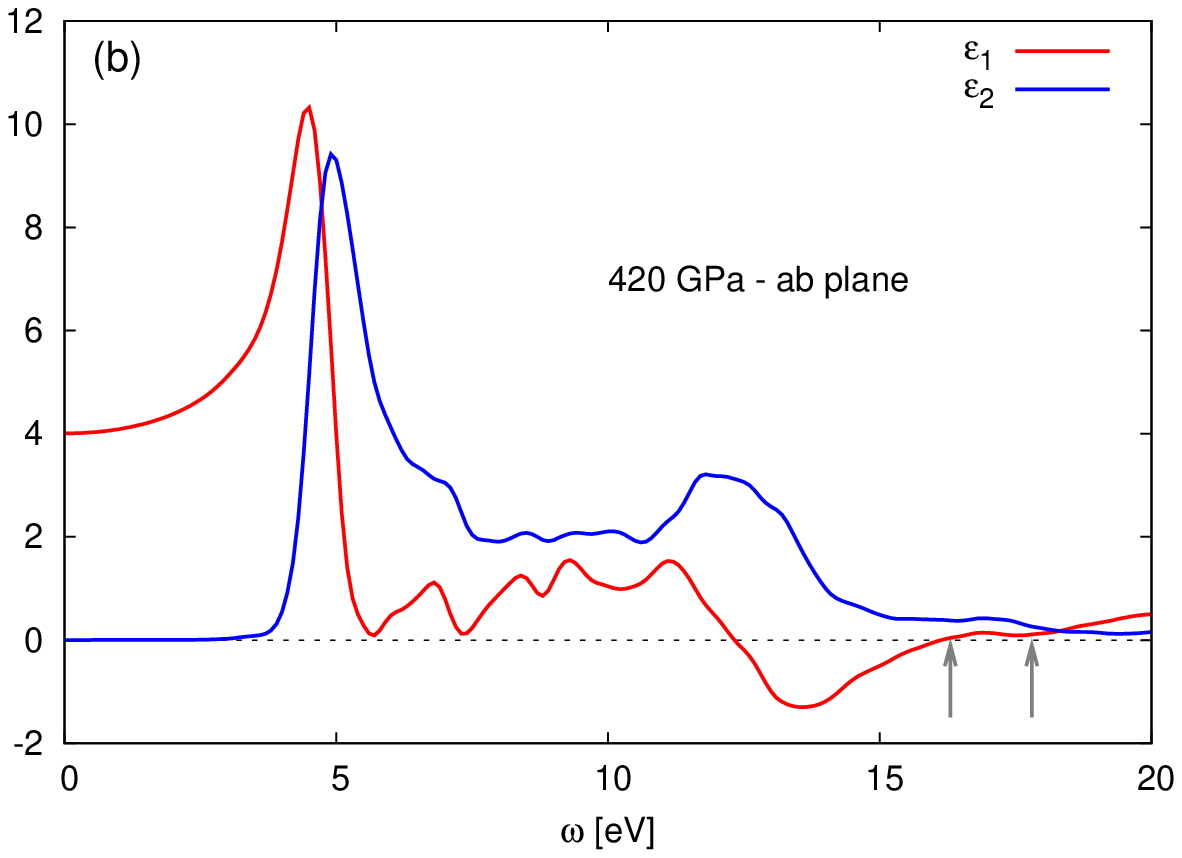} 
\includegraphics[width=0.49\columnwidth]{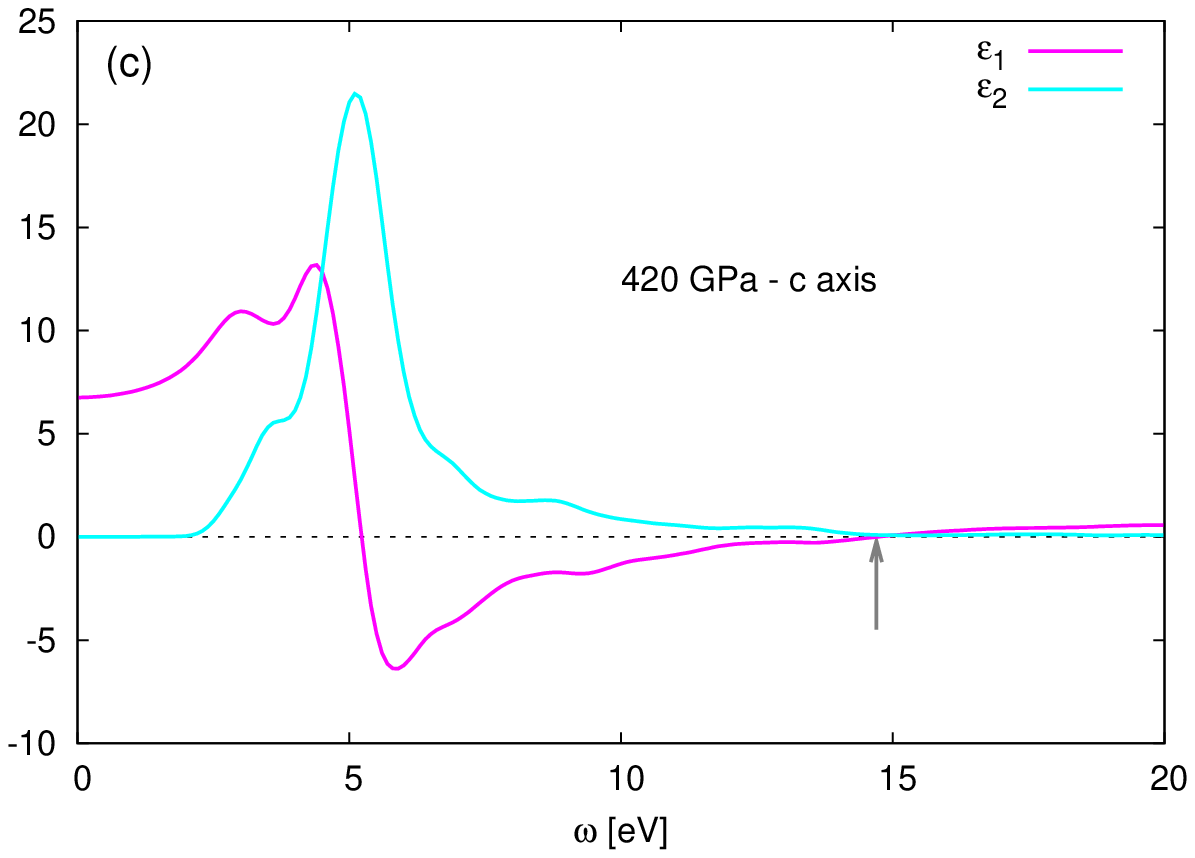}
\caption{{\bf Dielectric function.} 
Real and imaginary parts of the dielectric function $\epsilon = \epsilon_1 + i \epsilon_2$ for in-plane polarization (a) at 220 GPa and (b) 420 GPa, and $c$ axis polarization at 420 GPa (c) for a momentum transfer $\bfq=0.1$ reciprocal lattice units. The grey arrows mark the plasmon peaks in the energy loss spectra $L = \epsilon_2 / (\epsilon_1^2+\epsilon_2^2)$. They correspond to zeros of $\epsilon_1$ where also  $\epsilon_2$ is small. The first plasmon at 7 eV at the lowest pressure (a) is strongly suppressed by increasing the applied pressure: $\epsilon_2$ is rising in intensity in the same energy region (b), as an effect of band gap opening. The plasmon peak in the $c$ axis direction is located at lower energies. 
In fact. in the in-plane direction the main $\epsilon_2$ peak has a shoulder extending to higher energies (b) that is absent in the $c$ axis direction (c).}  
\label{figeps12}
\end{center}
\end{figure}

\end{widetext}


\begin{thebibliography}{99}

\bibitem{grosso} G. Grosso and G. Pastori-Parravicini, {\it Solid State Physics} (Academic, New York, 2000).

\bibitem{structure1} M. I. McMahon {\it et al.},  Proc. Natl. Acad. Sci. USA {\bf 105}, 17297 (2007).

\bibitem{structure2} E. Gregoryanz {\it et al.}, Science {\bf 320}, 1054 (2008).

\bibitem{melting1} E. Gregoryanz {\it et al.},  Phys. Rev. Lett. {\bf 94}, 185502 (2005).

\bibitem{nature} Y. Ma {\it et al.},  Nature  {\bf 458}, 182 (2009).

\bibitem{incommensurate} L. F. Lundegaard {\it et al.}, Phys. Rev. B {\bf 79}, 064105 (2009).

\bibitem{pnas}  A. Lazicki {\it et al.}, Proc. Natl. Acad. Sci. USA {\bf 106},  6525 (2009).
 
\bibitem{melting2} J.-Y. Raty, E. Schwegler, and S. A. Bonev, Nature {\bf 449},  448 (2007).


\bibitem{ashcroft-neaton} J. B. Neaton and N. W. Ashcroft, Phys. Rev. Lett. {\bf 86}, 2830 (2001).


\bibitem{mit2} N. E. Christensen, and D. L. Novikov,  Solid State Commun. {\bf 119}, 477 (2001).

\bibitem{ashcroft2008} B. Rousseau and N. W. Ashcroft, Phys. Rev. Lett. {\bf 101}, 046407 (2008).


\bibitem{ashcroft_comment} N. W. Ashcroft, Nature {\bf 458}, 158 (2009); Proc. Natl. Acad. Sci. USA {\bf 105}, 5 (2007).

\bibitem{RMP-noi} G. Onida, L. Reining, and A. Rubio, Rev. Mod. Phys. {\bf 74}, 601 (2002), and references therein.

\bibitem{Hedin} L. Hedin, Phys. Rev. {\bf 139}, A796 (1965).

\bibitem{epaps} See supplementary information. 

\bibitem{Godby-Needs}  R. W. Godby and  R. J. Needs,  Phys. Rev. Lett. {\bf 62}, 1169 (1989).

\bibitem{abbamonte} P. Abbamonte  {\it et al.}, 
 Proc. Natl. Acad. Sci. USA {\bf 105,} 12159 (2008).

\bibitem{dye} J. L. Dye,  Science {\bf 247,}  663 (1990).


\bibitem{adler-wiser} S. L. Adler,  Phys. Rev. {\bf 126}, 413 (1962);  N. Wiser,  Phys. Rev. {\bf 129}, 62 (1963).

\bibitem{hansi} H.-C. Weissker {\it et al.}, 
 Phys. Rev. Lett. {\bf 97,} 237602 (2006).

\bibitem{andrea} A. Marini, Phys. Rev. Lett. {\bf 101}, 106405 (2008). 

\bibitem{abinit} X. Gonze {\it et al.}, Z. Kristallogr. {\bf 220}, 558
(2005); see http://www.abinit.org.

\bibitem{dp-exc} The DP and EXC codes are developed by the French node of the ETSF; see
http://www.dp-code.org and http://www.bethe-salpeter.org.



\end{thebibliography}
\end{document}